\documentclass[11pt,amsmath,amssymb]{revtex4}
\usepackage{bm}
\begin{document}

\title{Unified approach to photo and electro-production \\
of mesons with arbitrary spins }
\author{G. Ramachandran}
 \affiliation{Indian Institute of Astrophysics,Koramangala,Bangalore,560034, India}
\author{M. S. Vidya}
 \affiliation{V/17, NCERT Campus, Sri Aurobindo Marg, New Delhi 110016, India}
\author{J. Balasubramayam}
 \altaffiliation{Department of Physics, Bangalore University, Bangalore, 560056, India} 
  \affiliation{K.S. Institute of Technology, Kanakapura Road,  Bangalore, 560062, India}
\date{\today}
\begin{abstract}
 A new approach to identify the independent amplitudes along with their 
partial wave multipole expansions, for photo and electro-production is suggested,
which is generally applicable to mesons with arbitrary spin-parity. These 
amplitudes facilitate direct identification of different resonance contributions.
\end{abstract}

\maketitle

\section{Introduction}

Improved experimental facilities to study photo and electro-production of 
pseudo-scalar and vector mesons have become available with the advent of the 
new generation of electron accelerators at JLab, MIT, BNL in USA, ELSA at Bonn,
MAMI at Mainz in Germany, ESRF at Grenoble in France and Spring8 at Osaka in Japan 
\cite{kru,bur}. With energies going up to 6 GeV, one can anticipate the extension 
of these studies to include mesons with higher spins $s > 1$, which are either
known already \cite{pdc} or have been predicted theoretically \cite{vij}. Since
$\eta, \omega, \phi $ are isoscalars in contrast to $\pi$ and $\rho$ which are isovectors, photo and 
electro-production of the former involve only the nucleon resonances in the intermediate state, 
whereas the latter involve the contributions from the delta resonances as well. Therefore these 
experimental studies assume importance in the context of the so called ``missing resonance problem" 
\cite{bar}, which is concerned 
with the resonances predicted on the basis of various theoretical models \cite{isg}
but have not been seen experimentally so far. In view of the dramatic violation 
of the OZI rule \cite{oku} observed in $\bar{p}p$ collisions \cite{ams}, and 
the measurement \cite{bal} of the ratio of $\phi / \omega$ production in 
$NN$ collisions, a similar   $\phi / \omega$  ratio  has been investigated recently
in the case of photo-production \cite{sib}. Photo-production of exotic mesons and
baryons have also attracted attention in recent years \cite{gui}.

Almost half a century ago a formalism for pion production was presented by 
Chew, Goldberger, Low and Nambu \cite{che} in the context of developing a 
dispersion theoretical approach to the problem. They expressed the photo pion production
amplitude in terms of four invariants $M_A, M_B, M_C, M_D$ and their coefficients 
$ A, B, C, D$ (which are essentially dependent on the c.m. energy $W$ and the angle $\theta$
in the c.m. frame between the meson and photon momenta denoted by $\boldsymbol{q}$ 
and ${\boldsymbol k}$ respectively). Using the two componental form for the nucleon 
spin the differential  cross-section for photo pion production was expressed 
in terms of four amplitudes denoted by ${\mathcal F}_1, {\mathcal F}_2, {\mathcal F}_3$
and ${\mathcal F}_4$ whose dependence on $\theta$ was made explicit through  
expansions involving the first and second derivatives of Legendre Polynomials 
with respect to $cos\theta$. The c.m. energy dependent `\emph{electric}' and 
`\emph{magnetic}' multipole partial wave amplitudes which appear theirin were 
denoted respectively by  $E_{l\pm}, M_{l\pm}$ where $l$ denote the relative angular 
momentum of the meson with  respect to nucleon in the final state and $\pm$ indicate 
the total angular momentum  $j=l\pm 1/2$. The authors themselves mentioned without giving 
any details that ``the derivation  of these formulae is lengthy  but can be 
carried out by straight forward methods". These formulae apply directly for  
photo production of other pseudo-scalar mesons like $\eta$. In the case  of the 
isovector pion, an earlier isospin analysis by Watson \cite{wat} was made use of 
to express each one of these amplitudes ${\mathcal F}_i, (i = 1-4)$ in
terms of three independent  nucleon isospin combinations ${\mathcal I}^{(+)}, \;
{\mathcal I}^{(-)},\;{\mathcal I}^{(0)}$,  multiplied respectively by 
${\mathcal F}_i^{(+)}, \;{\mathcal F}_i^{(-)}, \;{\mathcal F}_i^{(0)}$. Numerical 
estimates of these twelve amplitudes for pion production may be obtained, for 
example, from the second of the series of  three papers by Berends, Donnachie and 
Weaver \cite{ber} over a range of energies going up to 500 MeV. The first of the 
three papers \cite{ber} reviews the extension of the formalism \cite{che} to six 
amplitudes to include electro-production and describes also the connection
with helicity formalism \cite{jac}. The helicity approach was used recently \cite{pic}
to describe the photo production of vector mesons in terms twelve independent 
amplitudes. To our knowledge, no formalism exists as yet for electro-production 
of vector mesons or for photo and electro-production of tensor and higher spin mesons.

The purpose of the present paper is to suggest a simple and unified formalism for photo 
and electro-production of mesons with arbitrary spin-parity, $s^{\pi}$. It also has a built 
in isospin index $I$ along with the total angular momentum $j$, which makes 
it ideal for  readily identifying different baryon resonance contributions in the intermediate 
state.

\section{New amplitudes for photo and electro-production of mesons with arbitrary
spin-parity $s^{\pi}$}

Let us consider  photo production of a meson with spin-parity $s^{\pi}$ and isospin
$I_m$ at c.m. energy $W$. Let ${\boldsymbol k}$ and  ${\boldsymbol q}$ denote 
respectively the photon and meson momenta in c.m. frame. We use a  
right handed frame with z-axis chosen along ${\boldsymbol k}$ and the reaction plane 
containing ${\boldsymbol k}$ and  ${\boldsymbol q}$ as z-x plane. Using natural 
units with $\hbar, c$ and meson mass as unity, the  photon and meson energies 
in c.m. frame  are given in terms of $W$ by 
\begin{eqnarray}
k &=& \frac{1}{2W}[W^2-M^2] \\
\omega &=&  \frac{1}{2W}[W^2+1-M^2]= (q^2+1)^{{\textstyle{\frac{1}{2}}}},
\end{eqnarray}
where $M$ denotes the nucleon mass $k=|{\boldsymbol k}|$ and $q=|{\boldsymbol q}|$. 
The differential cross-section for the reaction in c.m. frame may then be written as 
\begin{equation}
\frac{d\sigma}{d\Omega} = \frac{q}{4k} \left(\frac{M}{4\pi W}\right)^2\sum_{ m_i m_f m_s \,p} 
|\langle 
{\textstyle{\frac{1}{2}}}\, m_f ;s\, m_s ; {\boldsymbol q}|T|
{\boldsymbol k},p\, ; {\textstyle{\frac{1}{2}}}\, m_i \rangle |^2,
\end{equation}
where  the initial and final nucleon spin projections are denoted by $m_i$ and $m_f$ 
respectively, the meson spin projection is denoted by $m_s$, while $p=\pm 1$ denote
left and right circular polarization states of photon as defined by \cite{ros}.
The covariant normalized $T$-matrix \cite{ber,don} is denoted by $T$. We may 
introduce the reaction amplitude ${\mathcal F}$ as in \cite{che}, through
\begin{equation}
\label{updcs}
\langle 
{\textstyle{\frac{1}{2}}}\, m_f ;s\, m_s ; {\boldsymbol q}|{\mathcal F}(p)|
{\boldsymbol k};{\textstyle{\frac{1}{2}}}\, m_i \rangle = \frac{M}{4\pi W}
\langle {\textstyle{\frac{1}{2}}}\, m_f ;s\, m_s ; {\boldsymbol q}|T|
{\boldsymbol k},p\, ; {\textstyle{\frac{1}{2}}}\, m_i \rangle.
\end{equation}

In the case of   electro-production, the momentum ${\boldsymbol k}$ of the 
virtual  photon is given by ${\boldsymbol k} = {\boldsymbol p}_i -{\boldsymbol p}_f$
 if  ${\boldsymbol p}_i $ and ${\boldsymbol p}_f $ denote the initial and final 
momenta of the electron.  The differential cross-section  $ d^5\sigma / d^3p_f d\Omega $ 
may be expressed following \cite{donb} in terms of the differential cross-section 
$d\sigma_v/ d\Omega $, for meson production by a virtual photon in the meson-nucleon 
c.m. frame. The amplitude for electro-production of mesons is similar to eq.\ref{updcs}, but derives 
contributions from longitudinal 
photons i.e., $p=0$ as well, in addition to $p=\pm 1$ 

Observing that the hadron spins  characterizing the entrance and exit channels
in the reaction are  respectively $s_i={\textstyle{\frac{1}{2}}}$
and $s_f$ takes the values $|s-{\textstyle{\frac{1}{2}}}|$ to $(s+{\textstyle
{\frac{1}{2}}})$, we may follow \cite{gr2} and using the same notation introduce 
operators 
\begin{equation}
S^{\lambda}_{\mu}(s_f,{\textstyle\frac{1}{2}})= \sum_{n=0}^1 \frac{[s_f]^2\,[n]}{\sqrt{2}\,[s]}
W(\lambda {\textstyle\frac{1}{2}} s {\textstyle\frac{1}{2}}; s_f\,n) 
\left(S^{s}(s,0) \otimes S^{n
}({\textstyle\frac{1}{2}}, {\textstyle\frac{1}{2}})\right)^{\lambda }_{\mu},
\end{equation}
which are irreducible tensor operators of rank $\lambda=|s_f-{\textstyle\frac{1}{2}}|$
to $(s_f+{\textstyle\frac{1}{2}})$ in hadron spin space and 
express  in the reaction amplitude ${\cal F}(p)$ given by eq.\ref{updcs} as 
\begin{equation}
{\cal F}(p)= \sum_{\lambda = |s-n|}^{(s+n)} \sum_{n=0}^1 \left((S^{s}(s,0) \otimes S^{n
}({\textstyle\frac{1}{2}}, {\textstyle\frac{1}{2}}))^{\lambda }
\cdot {\cal F}^{\lambda}(n,p)\right),
\label{amp}
\end{equation}
in terms of irreducible tensor amplitudes ${\cal F}^{\lambda}_{\mu}(n,p)$, which
constitute the new basic amplitudes in our formalism.

To obtain formulae for these new amplitudes in terms of the different partial
wave multipole amplitudes, we express $\langle {\boldsymbol q}|$ and $|{\boldsymbol k},p \rangle$ in 
right hand side of eq.\ref{updcs} in terms of partial waves and  multipoles \cite{ros} respectively, 
using 
\begin{eqnarray}
e^{-i\boldsymbol{q \cdot r}}&=& 4\pi \sum_{l=0}^{\infty} (-i)^l j_{l}
(qr)\sum_{m_l=-l}^{l}Y_{lm_l}(\hat{{\boldsymbol q}}) Y_{lm_l}(\hat{{\boldsymbol r}})^*,
\label{pwe} \\
\label{fpwe}
\hat{{\boldsymbol u}}_p\, e^{i\boldsymbol{k \cdot r}} &=& \sqrt{2\pi}\sum_{L=|p|}^{\infty}i^L [L]
\left[|p|{\boldsymbol A}^{(m)}_{Lp}({\boldsymbol r}) + ip\,  {\boldsymbol A}^{(e)}_{Lp}({\boldsymbol 
r})]- i \sqrt{2}\,
(1-p^2){\boldsymbol A}^{(\ell)}_{Lp}({\boldsymbol r})\right],
\end{eqnarray}
where the \emph{`magnetic' , `electric'} and \emph{`longitudinal'} $2^L$ poles states 
of the photon are given  respectively by
\begin{eqnarray}
{\boldsymbol A}^{(m)}_{Lp}({\boldsymbol r}) &=& j_L(kr)\,{\boldsymbol T}_{LLp}(\hat{{\boldsymbol 
r}}) \\
\label{emp}
{\boldsymbol A}^{(e)}_{Lp}({\boldsymbol r}) &=& -\sqrt{\frac{L}{2L+1}}\,j_{L+1}(kr){\boldsymbol 
T}_{LL+1p}(\hat{{\boldsymbol r}})
+\sqrt{\frac{L+1}{2L+1}}\,j_{L-1}(kr){\boldsymbol T}_{LL-1p}(\hat{{\boldsymbol r}})\\
\label{lmp}
{\boldsymbol A}^{(\ell)}_{Lp}({\boldsymbol r}) &=& \sqrt{\frac{L+1}{2L+1}}\,j_{L+1}(kr){\boldsymbol 
T}_{LL+1p}(\hat{{\boldsymbol r}})
+\sqrt{\frac{L}{2L+1}}\,j_{L-1}(kr){\boldsymbol T}_{LL-1p}(\hat{{\boldsymbol r}}),
\end{eqnarray}
in terms of the vector spherical harmonics
\begin{equation}
{\boldsymbol T}_{Ll_{\gamma}M}(\hat{{\boldsymbol r}}) = \sum_p C(l_{\gamma}1L;m_{\gamma}pM)
Y_{l_{\gamma}m_{\gamma}}(\hat{{\boldsymbol r}}) \hat{{\boldsymbol \xi}}_p,
\end{equation}
which are irreducible tensors of rank $L$. When the z-axis is chosen along 
${\boldsymbol k}$, the summation over $p$ drops to a single term with $M=p$ since
$m_{\gamma}$ can assume only one value  $m_{\gamma}=0$. The 
quantum numbers $l_{\gamma}$ and $L$ correspond respectively to the orbital and
total angular momenta of the photon.  If $\hat{{\boldsymbol e}}_1, \hat{{\boldsymbol e}}_2,
\hat{{\boldsymbol e}}_3$ denote mutually orthogonal unit vectors constituting a right 
handed system such that $\hat{{\boldsymbol e}}_3$ along ${\boldsymbol k}$, then $\hat{{\boldsymbol 
\xi}}_
{\pm 1}, \; \hat{{\boldsymbol \xi}}_0$ are defined by 
\begin{equation}
\hat{{\boldsymbol \xi}}_{\pm 1} = \mp \frac{1}{\sqrt{2}}(\hat{{\boldsymbol e}}_1 \pm i\,
\hat{{\boldsymbol e}}_2)\;\; , \;\; \hat{{\boldsymbol \xi}}_0= \hat{{\boldsymbol e}}_3,
\end{equation}
and $\hat{{\boldsymbol u}}_p$ in eq.\ref{fpwe} are given by $\hat{{\boldsymbol u}}_p = 
-p \, \hat{{\boldsymbol \xi}}_p , \; p=\pm 1 \; ; \; \hat{{\boldsymbol u}}_0 = 
\hat{{\boldsymbol \xi}}_0$, where $\hat{{\boldsymbol u}}_{\pm 1}$ correspond to left and 
right circular polarization states as defined by Rose \cite{ros}.

Combining the total angular momentum $L$ of the photon with the initial nucleon 
spin ${\textstyle\frac{1}{2}}$ to yield the total angular momentum $j$ which is 
conserved in the reaction, it is clear that the  same $j$ is obtained in combining 
the orbital angular momentum $l$ of the meson with channel spin $s_f$ in the final 
state. We thus have 
\begin{eqnarray}
\label{famp}
\langle {\textstyle{\frac{1}{2}}}\, m_f ;s\, m_s ; {\boldsymbol q}|{\mathcal F}(p)|
{\boldsymbol k} ; {\textstyle{\frac{1}{2}}}\, m_i \rangle &=& 
 \frac{(2\pi)^{{\textstyle\frac{1}{2}}}M}{W}\sum_{l=0}^{\infty}
\sum_{s_f=|s-{\textstyle{\frac{1}{2}}}|}^{(s+{\textstyle{\frac{1}{2}}})}
\sum_{L=|p|}^{\infty} \sum_{j=L-{\textstyle{\frac{1}{2}}}}^{L+{\textstyle{\frac{1}{2}}}} 
\,  i^{L-l} [L] \nonumber \\ & & \times  \langle(l(s{\textstyle\frac{1}{2}})s_f)j\|T\|
(L{\textstyle\frac{1}{2}})j\rangle
C(s {\textstyle{\frac{1}{2}}} s_f ; m_s m_f m_{s_f}) \nonumber \\ & & \times  
C(l s_f j ; m_l m_{s_f} m)\, C(L {\textstyle{\frac{1}{2}}} j ; p m_i m)\, Y_{l m_l}(\theta, 0),
\end{eqnarray}
where the reduced matrix elements 
depend only on the c.m. energy $W$ and the angular dependence 
is completely taken care of by $Y_{l m_l}(\theta, 0)$ where $m_l=p+m_i-m_f-m_s$.  
We may express 
\begin{eqnarray}
C(L {\textstyle{\frac{1}{2}}} j ; p m_i m)C(l s_f j ; m_l m_{s_f} m) &=&
\sum_{\lambda} W(L{\textstyle\frac{1}{2}}ls_f;j\lambda)  [j]^2[\lambda][s_f]^{-1} 
(-1)^{l+L+{\textstyle\frac{1}{2}}-j} \nonumber \\ & & \times (-1)^{p+\mu} 
C({\textstyle\frac{1}{2}}\lambda s_f;m_i-\mu m_{s_f})\nonumber \\ & & \times
C(lL\lambda;m_l-p\mu),
\end{eqnarray} 
and replace 
\begin{eqnarray}
C(s {\textstyle{\frac{1}{2}}} s_f ; m_s m_f m_{s_f})C({\textstyle\frac{1}{2}}
\lambda s_f;m_i-\mu m_{s_f})[\lambda] &=& S_{-\mu}^{\lambda}(s_f,
{\textstyle{\frac{1}{2}}})   =  \sum_{n=0}^1 \frac{[s_f]^2\,[n]}{\sqrt{2}\,[s]}
W(\lambda {\textstyle\frac{1}{2}} s {\textstyle\frac{1}{2}}; s_f\,n)
\nonumber \\ & & \times \left(S^{s}(s,0) \otimes S^{n}({\textstyle\frac{1}{2}}, 
{\textstyle\frac{1}{2}})\right)^{\lambda }_{-\mu},
\end{eqnarray} 
so that we have a single elegant formula viz.
\begin{eqnarray}
\label{samp}
{\cal F}_{\mu}^{\lambda}(n,p)
&=&  [n]\sum_{l=0}^{\infty}
\sum_{s_f=|s-{\textstyle{\frac{1}{2}}}|}^{(s+{\textstyle{\frac{1}{2}}})}
\sum_{L=|p|}^{\infty} \sum_{j=L-{\textstyle{\frac{1}{2}}}}^{L+{\textstyle{\frac{1}{2}}}}  
\nonumber \\ & & \times W(\lambda {\textstyle\frac{1}{2}} s {\textstyle\frac{1}{2}}; s_f\,n)
W(L{\textstyle\frac{1}{2}}ls_f;j\lambda) {\cal F}_{ls_f;L}^{j}
{\cal A}_{\mu}^{\lambda}(\theta),
\end{eqnarray}
for expressing the irreducible tensor amplitudes ${\cal F}_{\mu}^{\lambda}(n,p)$
for all allowed values of $\lambda, \mu$ in terms of the partial wave multipole 
amplitudes 
\begin{equation}
{\cal F}_{ls_f;L}^{j} =  \frac{\sqrt{\pi}M}{W}
i^{l+L}  (-1)^{L+{{\scriptstyle\frac{1}{2}}}-j}[L][j]^2[s_f][s]^{-1}
\langle(l(s{\textstyle\frac{1}{2}})s_f)j\|T\|(L{\textstyle\frac{1}{2}})j\rangle
\end{equation}
for photo or electro-production of mesons. The ${\cal F}_{ls_f;L}^{j}$  
may explicitly be written either as `\emph{magnetic}'  or `\emph{electric}' or 
`\emph{longitudinal}' using eq.\ref{fpwe} and taking parity conservation into 
account.We have
\begin{equation}
\label{fpwe2}
{\cal F}_{ls_f;L}^{j} = |p|f_-{\cal M}_{ls_f;L}^{j} + i p f_+ 
{\cal E}_{ls_f;L}^{j}-i\sqrt{2}(1-p^2 )f_+{\cal L}^{j}_{ls_f;L},
\end{equation}
in terms of the `\emph{electric}',  `\emph{magnetic}' and `\emph{longitudinal}' 
multipole partial wave amplitudes denoted respectively by ${\mathcal E}_{ls_f;L}^{j}, 
\; {\mathcal M}_{ls_f;L}^{j}$ and ${\mathcal L}_{ls_f;L}^{j}$.
  The factors 
\begin{equation}
f_{\pm} = \frac{1}{2}[1\pm \pi (-1)^{L+l}]
\end{equation}
assume values either 1 or 0 to ensure parity conservation.  The angular dependence of 
${\cal F}_{\mu}^{\lambda}(n,p)$ is fully contained in 
\begin{equation}
\label{alm}
{\cal A}_{\mu}^{\lambda}(\theta)=
(-1)^{p}C(lL\lambda;m_l-p\mu)Y_{lm_l}(\theta,0),
\end{equation}
which are independent of $s_f, j$ and isospin quantum numbers.
Isospin considerations lead to 
\begin{equation}
\label{iso}
{\mathcal F}_{l s_f ; L}^j = \sum_{ I_{\gamma}}\sum_{I=|I_m-{\textstyle{\frac{1}{2}}}|}^
{(I_m+{\textstyle{\frac{1}{2}}})}C({\textstyle{\frac{1}{2}}} I_m I ; 
\nu_f \nu_m \nu_i)\,C({\textstyle{\frac{1}{2}}} I_{\gamma} I; \nu_i 0 \nu_i)\,
{\mathcal F}_{l s_f ; L}^{ I_{\gamma}I j},
\end{equation}
where $\nu_i$ and $\nu_f$ denote the nucleon isospin projections in the initial 
and final states respectively, while $\nu_m$ denotes the meson isospin  projection. 
The isospin $I_{\gamma}$ of the photon \cite{pdc} takes the values $0,1$. It was 
suggested  \cite{san} that there could also be an isotensor component $I_{\gamma}=2$ 
for the photon, in which case the summation over $I_{\gamma}$ may be extended 
from 0 to 2 and one can look out in experiments for non-zero ${\mathcal F}_
{l s_f ; L}^{ I_{\gamma}I j}$  with $I_{\gamma}=2$. It is important to note  that 
the superscripts $I,j$ of ${\mathcal F}_{l s_f ; L}^{ I_{\gamma}I j}$ in our 
formalism may readily be identified with the isospin and spin quantum numbers 
$I, j$ of the contributing resonances in the intermediate state. It may also be noted 
that the precise composition of ${\mathcal F}_{l s_f ; L}^j$ in terms of 
${\mathcal F}_{l s_f ; L}^{ I_{\gamma}I j}$ is known using  eq.\ref{iso}, when 
the charge states of the hadrons are specified and that  the isospin indices 
$I_{\gamma}, I$ are to be attached not only to the amplitudes  on the left hand 
side but also exactly identically to those on the right hand side of eq.\ref{fpwe2}. 

It is important to note that the  ${\cal F}_{\mu}^{\lambda}(n,p)$ satisfy the 
symmetry property 
\begin{equation}
\label{sym}
{\cal F}_{-\mu}^{\lambda}(n ,-p)=
\pi(-1)^{\lambda-\mu}{\cal F}_{\mu}^{\lambda}(n,p),
\end{equation}
which enables us to determine the number of independent amplitudes in any given 
case. It may also be noted that $S^0_0({\textstyle\frac{1}{2}},{\textstyle\frac{1}{2}})$
is a unit $2\times 2$ matrix and 
\begin{equation}
S^1_{\pm 1}({\textstyle\frac{1}{2}},{\textstyle\frac{1}{2}})= \mp \frac{1}{\sqrt{2}}
(\sigma_x \pm i\,\sigma_y)\; ; \; S^1_0({\textstyle\frac{1}{2}},{\textstyle\frac{1}{2}})=
\sigma_z,
\end{equation}
where $\sigma_x,\,\sigma_y,\,\sigma_z$ denote the Pauli spin matrices for the 
nucleon.

\section{Particular Cases}
 \subsection{Photo and electro-production of pseudo-scalar mesons $(s^{\pi}=0^-)$}
 
The new basic amplitudes in this well-known case  are ${\mathcal F}^0_0(0,p),
{\mathcal F}^1_{\mu}(1,p)$ with $\mu = \pm 1 ,0$ and $p=\pm 1$,  since $s=0$ 
implies $\lambda = n = 0,1$. The use of  eq.\ref{sym} implies 
\begin{eqnarray}
\label{sfoo}
{\cal F}_{0}^{0}(0,1) &=& -{\cal F}_{0}^{0}(0,-1)\\
{\cal F}_{0}^{1}(1,1) &=&+{\cal F}_{0}^{1}(1,-1) \\ 
{\cal F}_{1}^{1}(1,1) &=&-{\cal F}_{-1}^{1}(1,-1)\\
{\cal F}_{-1}^{1}(1,1) &=&-{\cal F}_{+1}^{1}(1,-1)
\end{eqnarray}

Thus, the number of independent amplitudes is the same as in \cite{che} i.e., four.
We note moreover that only one   amplitude viz. ${\cal F}_{0}^{0}(0,1)$ 
is spin independent, whereas all the other three are spin dependent, which is 
the case in \cite{che} as well.

In the case of electro-production, we have two additional independent basic 
amplitudes viz., 
\begin{eqnarray}
\label{sf10}
{\cal F}_{0}^{1}(1,0) & & \\
\label{sf11}
{\cal F}_{+1}^{1}(1,0) &=& -{\cal F}_{-1}^{1}(1,0)
\end{eqnarray}
thus making the total six , which is consistent with \cite{ber}. The two additional
amplitudes ${\cal F}_{0}^{1}(1,0)$ and ${\cal F}_{+1}^{1}(1,0)$ are spin 
dependent as in \cite{ber}.

\subsection{Photo and electro-production of vector mesons $(s^{\pi}=1^-)$}
The independent basic amplitudes for vector meson photo-production in our 
formalism are 
\begin{eqnarray}
{\cal F}_{0}^{1}(0,1)=+{\cal F}_{0}^{1}(0,-1)\\
{\cal F}_{+1}^{1}(0,1)=-{\cal F}_{-1}^{1}(0,-1)\\
{\cal F}_{-1}^{1}(0,1)=-{\cal F}_{+1}^{1}(0,-1)
\end{eqnarray}
which are spin independent and 
\begin{eqnarray}
{\cal F}_{0}^{0}(1,1)=-{\cal F}_{0}^{0}(1,-1)\\
{\cal F}_{0}^{1}(1,1)=+{\cal F}_{0}^{1}(1,-1) \\
{\cal F}_{+1}^{1}(1,1)=-{\cal F}_{-1}^{1}(1,-1)\\
{\cal F}_{-1}^{1}(1,1)=-{\cal F}_{+1}^{1}(1,-1)\\
{\cal F}_{0}^{2}(1,1)=-{\cal F}_{0}^{2}(1,-1) \\
{\cal F}_{+1}^{2}(1,1)=+{\cal F}_{-1}^{2}(1,-1)\\
{\cal F}_{-1}^{2}(1,1)=+{\cal F}_{+1}^{2}(1,-1)\\
{\cal F}_{+2}^{2}(1,1)=-{\cal F}_{-2}^{2}(1,-1)\\
{\cal F}_{-2}^{2}(1,1)=-{\cal F}_{+2}^{2}(1,-1)
\end{eqnarray}
which are spin dependent, taking the total to twelve independent amplitudes, which is
in agreement with \cite{pic}.

Electro-production  has not been considered either in \cite{pic} or by others. In our 
formalism, we readily identify the additional independent amplitudes as 
\begin{eqnarray}
\label{ve10}
{\cal F}_{0}^{1}(0,0) & & \\
\label{ve11}
{\cal F}_{1}^{1}(0,0) &=& -{\cal F}_{-1}^{1}(0,0)\\
{\cal F}_{0}^{1}(1,0) & & \\
{\cal F}_{1}^{1}(1,0) &=&-{\cal F}_{-1}^{1}(1,0) \\
{\cal F}_{1}^{2}(1,0)&=&+{\cal F}_{-1}^{2}(1,0) \\
{\cal F}_{2}^{2}(1,0)&=&-{\cal F}_{-2}^{2}(1,0),
\end{eqnarray}
i.e., six in addition to twelve taking the total to eighteen. Of the additional 
six, two  amplitudes ${\cal F}_{0}^{1}(0,0)$ and ${\cal F}_{1}^{1}(0,0)$ are 
spin independent, and the remaining are spin dependent. 

\subsection{Photo and electro-production of tensor mesons $(s^{\pi}=2^+)$}

Application of the symmetry relation represented by eq.\ref{sym} shows that 
photo-production of tensor mesons with spin parity $s^{\pi}=2^+$ is 
characterized by a set of twenty independent irreducible
tensor amplitudes whereas electro-production of tensor mesons needs an additional 
ten amplitudes, thus taking the total to thirty.
 
\section{Summary and outlook}
A theoretical formalism has been outlined in this paper for photo and electro-production of mesons 
with arbitrary spin-parity $s^{\pi}$, where the reaction 
amplitude $ {\mathcal F}$ in each case is expressed in terms of a basic set of 
independent irreducible tensor amplitudes $ {\mathcal F}^{\lambda}_{\mu}(n,p)$
of rank $\lambda$. The number of independent amplitudes in our formalism is in
agreement with the number determined in some particular cases employing different 
arguments by earlier authors.  For example, the number of independent irreducible
tensor amplitude is four  in the case of photo-production of pseudo-scalar mesons. 
The four different independent amplitudes have been introduced  in a different 
way by Chew, Goldberger, Low and Nambu in their famous paper \cite{che}. Each of 
the four amplitudes of CGLN have different formulae for expressing them  in terms 
of partial wave multipole amplitudes. A highlight of our approach is that a single 
elegant formula namely eq.\ref{samp} describes the expansion of the independent 
irreducible tensor amplitudes in terms of the partial wave multipole amplitudes, 
irrespective of whether they are four as in the case of photo-production of 
pseudo-scalar mesons or six in the case of  electro-production of pseudo-scalar 
mesons or ten in the case of photo-production of vector mesons or eighteen as in 
the case of electro-production  of vector mesons or twenty in the case of 
photo-production of tensor mesons or thirty in the case of electro-production of 
tensor mesons. For the photo-production of isovector mesons, CGLN employ Watson's 
approach for isospin indexing of each of their amplitudes by three superscripts 
$(+), (-), (0)$,  specific linear combinations of which have to be taken for the 
reaction when initial and final charge states are given. Instead  our amplitudes 
carry a specific isospin index $I$. The explicit  superscripts $I,j$  permit us 
to identify directly the resonance contributions coming from the intermediate 
states.
 
As more and more experimental data are forthcoming at higher energies, we hope 
that the unified and simpler formalism outlined in this paper will be found 
useful to  analyze  measurements.  We have not 
explicitly written down the new basic amplitudes of our formalism for mesons 
with spin $s \geq 2$, as experimental studies are yet to be reported . 
However,it is clear that 
the formalism is readily extendable to photo and electro-production of mesons 
like $f_2(1270)$ or $a_2(1320)$ or $f_2'(1525)$ with spin parity $2^+$ or
$\pi_2(1670)$ with spin-parity $2^-$ and $\omega_3(1670)$ or $f_3(1690)$  with 
spin-parity $3^-$ and $a_4(2040)$ or $f_4(2050)$ with spin-parity $4^+$ which
are known \cite{pdc} to exist. Since the energy at J.Lab can go up to 6 GeV, it 
is clearly possible to reach the thresholds for production of these higher spin 
mesons.

\begin{acknowledgments}
One of us (G.R.) is grateful to Professors B.V. Sreekantan, R. Cowsik, J.H. Sastry, 
R. Srinivasan and S.S. Hassan for facilities provided for research at the  
Indian Institute of Astrophysics and another (J.B.) acknowledges much encouragement 
for research given by the Principal Dr. T.G.S. Moorthy and the Management of 
K.S. Institute of Technology.
\end{acknowledgments}

\end{document}